\documentclass{revtex4-1}
\draft
\usepackage{graphicx}% Include figure files
\usepackage[calcwidth = .9\linewidth, font = small, labelfont = bf]{caption}
\usepackage{dcolumn}% Align table columns on decimal point
\usepackage{bm}% bold math
\usepackage{xcolor}
\usepackage[normalem]{ulem}

\usepackage[utf8]{inputenc}
\usepackage[T1]{fontenc}
\usepackage{mathptmx}
\usepackage{etoolbox}
\usepackage{amsmath}
\usepackage{relsize}
\usepackage{float}
\usepackage{xfrac}
\usepackage[margin=2cm]{geometry}
\usepackage{adjustbox}
\usepackage{multirow}
\usepackage{xcolor, soul}
\usepackage[normalem]{ulem}
\sethlcolor{green}

\usepackage{hyperref}

\makeatletter
\def\@email#1#2{%
 \endgroup
 \patchcmd{\titleblock@produce}
  {\frontmatter@RRAPformat}
  {\frontmatter@RRAPformat{\produce@RRAP{*#1\href{mailto:#2}{#2}}}\frontmatter@RRAPformat}
  {}{}
}%
\makeatother

\newcommand{\ltx}{LTX-$\beta$~}

\newcommand{\tps}{$\tau_p^*$~}

\begin{document}

\preprint{AIP/123-QED}

\title[A plasma-material interaction design basis for NSTX-U flash lithium evaporators]{A plasma-material interaction design basis for NSTX-U flash lithium evaporators}
%Addressing PMI requirements with flash lithium evaporator designs/development in \ltx and NSTX-U
%to address PMI needs 
%Liquid Li evaporator refill in \ltx and design of flash Li evaporators for NSTX-U 
%wall conditioning
%based on \ltx designs
% Force line breaks with \\
%\author{A. Maan}

% \altaffiliation[Also at ]{Physics Departmßent, XYZ University.}%Lines break automatically or can be forced with \\
% \email{Second.Author@institution.edu.}
%\affiliation{ 
%PPPL%\\This line break forced with \textbackslash\textbackslash
%}%
\author{A. Maan, R. Majeski, C. López Pérez, D. P. Boyle, T. Le, R. Lunsford}
\affiliation{%
Princeton Plasma Physics Laboratory, Princeton, NJ, USA
}%
\homepage{Corresponding author: amaan@pppl.gov}
\date{\today}% It is always \today, today,

\begin{abstract}

Plasma-material interaction (PMI) at lithium conditioned plasma facing components (PFCs) depends not only on lithium inventory, but also on coating freshness, chemical state, and spatial coverage. We use LTX-$\beta$ measurements and lithium evaporator development to translate these constraints into a design basis for the NSTX-U flash lithium evaporator (f-LITER). During an LTX-$\beta$ run day with evaporations at the beginning and midpoint of the run day, discharges immediately after each evaporation exhibited higher plasma current, lower line averaged density, and shorter effective particle confinement times than later discharges with identical programmed fueling and coil currents, consistent with stronger particle pumping by the freshly conditioned wall. The subsequent relaxation is interpreted in the context of published in vacuo x-ray photoelectron spectroscopy observations of Li$_2$O growth and molecular dynamics calculations showing that deuterium reflection and retention depend on oxide thickness. To refresh this evolving interface while limiting contamination, a low thermal mass evaporator was coupled to an in vacuo liquid lithium dropper and tested on LTX-$\beta$ and in a dedicated test chamber. Quartz crystal microbalance measurements gave a 401 nm deposition in 17.4 min, equivalent to 100 nm in approximately 4.8 min of active evaporation. A qualitative temperature programmed desorption comparison showed weaker impurity desorption from a coating deposited by the evaporator loaded using the liquid dropper than from one deposited by the evaporator loaded with solid lithium. These results establish the PMI requirements for an NSTX-U f-LITER architecture based on the LTX-$\beta$ low thermal mass evaporators. The system is intended to make lithium delivery a reproducible PMI actuator for controlling recycling, impurity uptake, and PFC conditioning in double null NSTX-U operation.
\end{abstract}

\maketitle

\section{\label{sec:intro}Introduction}

Lithium wall conditioning can alter the plasma boundary by reducing hydrogenic recycling and chemically binding impurities – thereby modifying the conditions that set sources, profiles, stability, transport, and radiative losses in the plasma edge. As a candidate liquid metal for a future Fusion Pilot Plant (FPP) or Liquid Metal Core-Edge integration facility (LMCE), lithium could also offer new ways to deal with extreme heat and particle fluxes via flow and/or evaporation \cite{Rindt_2021,Ruzic_2011,FISHER2020100855,Menard_LMCE_whitepaper}. The low atomic number of lithium makes it a comparatively benign plasma impurity, while its chemical reactivity with hydrogen, oxygen, carbon, and other residual species can reduce neutral fueling from the wall and suppress impurity influx from plasma facing components (PFCs). 

%general li stuff not just low r, related to other effects including potentially high r vapor stuff.

Reducing the fraction of escaping ions that recycle as neutrals can raise edge temperatures, reduce edge collisionality, and decrease temperature gradients that drive turbulent transport \cite{maan2024estimates,majeski_pop,ZAKHAROV2004149}. The Lithium Tokamak Experiment (LTX) and its successor \ltx provided a direct experimental test of the low recycling boundary condition by operating with lithium coated PFCs. In LTX, lithium conditioning enabled discharges with reduced gradient or nearly flat electron temperature profiles, behavior consistent with reduced edge neutral cooling and reduced recycling at the plasma boundary \cite{boyle-prl,majeski_pop}. These discharges also showed low scrape-off-layer density and low impurity content in the core despite the elevated plasma temperatures, indicating that lithium conditioned walls could simultaneously support hot edge operation and acceptable impurity levels. 

Separate \ltx measurements showed that lithium coatings improved plasma and neutral density control: plasma current, discharge duration, edge temperature, and density pumpout increased while plasma density and neutral influx decreased \cite{MAAN2023_NME}. Estimates of the global recycling coefficient further showed that the effective particle confinement time can approach the electron energy confinement time in the lowest recycling discharges \cite{maan2024estimates}. Solid lithium coatings, however, slowly oxidized to Li$_2$O from the top at the timescale of hours. The chemical species mix and elemental composition of the PFCs were diagnosed with in vacuo surface science measurements and showed that modest Li$_2$O concentration did not dramatically degrade performance. However, when the oxide growth transitioned to LiOH over time, a step change in performance degradation was observed. Together with surface analysis showing oxidized but still functionally active lithium coatings on \ltx PFCs \cite{Maan_PPCF_2020,Maan_IEEE_2020}, these results make the freshness, coverage, and repeatability of lithium delivery central engineering requirements rather than secondary operational details.

The National Spherical Torus Experiment (NSTX) previously used evaporative lithium coatings to condition graphite PFCs, and measurements with progressively increasing lithium deposition showed reduced recycling, modified edge profiles, edge localized mode (ELM) suppression, and improved confinement trends relative to boronized reference conditions \cite{NSTX,Maingi_2012}. NSTX liquid lithium divertor experiments further indicated that plasma response depended not only on the amount of lithium introduced, but also on the condition and chemical state (i.e. metallic, oxidized or carbonated lithium) of the exposed lithium surface \cite{Ono_2017}. These NSTX results provide the motivation for applying the LTX and \ltx evaporator lessons to NSTX-U, where lithium delivery must be compatible with the larger device geometry, integration constraints, and the need for reproducible wall conditioning.

The remainder of the paper is organized as follows. Section \ref{sec:shots} first establishes the plasma-material interaction requirements using a new analysis of shot to shot evolution during an \ltx run day with lithium evaporations at the beginning and midpoint of the run day, interpreted in the context of lithium surface oxidation. Section \ref{sec:evolution} then presents the LTX and \ltx evaporator design evolution, relating changes in coating coverage, wetting behavior, inventory, impurity control, and serviceability to those PMI requirements. Section \ref{sec:mark2} discusses limitations of the legacy NSTX evaporators and describes the present flash lithium evaporator (f-LITER) concept, including the transition from a manually loaded solid lithium source to an evaporator loaded in vacuo using the liquid dropper. Temperature programmed desorption (TPD) of a witness sample is used as a qualitative comparison of impurity bearing desorption following the two loading methods. The proposed NSTX-U f-LITER design incorporating these refinements is then presented, and Section \ref{sec:discussion} summarizes the resulting design basis.

%leading up to the operation and refill of an evaporator,
This paper documents the design evolution of \ltx lithium evaporators to optimize the plasma-material interface, forming an engineering design basis for NSTX-U. The principal contribution is the translation of LTX and \ltx operating experience into NSTX-U evaporator requirements, followed by demonstration in \ltx of a flash evaporator that addresses those requirements, and design plans for NSTX-U. We show that wall freshness, spatial coverage, and repeatable, reproducible lithium deposition are critical design requirements that follow from the observed sensitivity of plasma performance to lithium coating history and surface chemical evolution \cite{Maan_PPCF_2020}, rather than separate hardware preferences. The proposed NSTX-U design will feature a liquid lithium dropper to load the evaporator in vacuo in order to limit impurity uptake during loading. %Demonstration of in vacuo loading of a lithium evaporator 
%The proposed NSTX-U design will feature a liquid lithium dropper to load the evaporator in vacuo in order to limit impurity uptake during loading.160

%We show that wall freshness, spatial coverage, and repeatable lithium inventory are design requirements for a successful lithium evaporator implementation that follow from the observed sensitivity of plasma performance to lithium coating  evolution \cite{Maan_PPCF_2020}, rather than separate hardware preferences.
%documents the in vacuo liquid Li loading and operation of a lithium evaporator.
%and the proposed NSTX-U design, which features a liquid lithium dropper to load the evaporator in vacuo in order to limit impurity uptake during loading.
\section{\label{sec:shots}Plasma-material interaction requirements from between shot lithium evaporation in \ltx}

The plasma response to lithium conditioning provides the physical basis for the evaporator requirements developed in this paper. Previous \ltx studies compared run day ensemble averages under wall conditions established by a single evaporation at the beginning of the run day. NSTX-U, however, requires the ability to refresh lithium during a run day. This section therefore examines the shot to shot evolution of plasma and surface parameters during an \ltx experiment that included a second lithium evaporation midway through the run day. The analysis tests whether the wall pumping state established by evaporation remains constant or evolves on an operationally relevant timescale.

\begin{figure}[htbp]
\centering
\includegraphics[width=0.45\columnwidth]{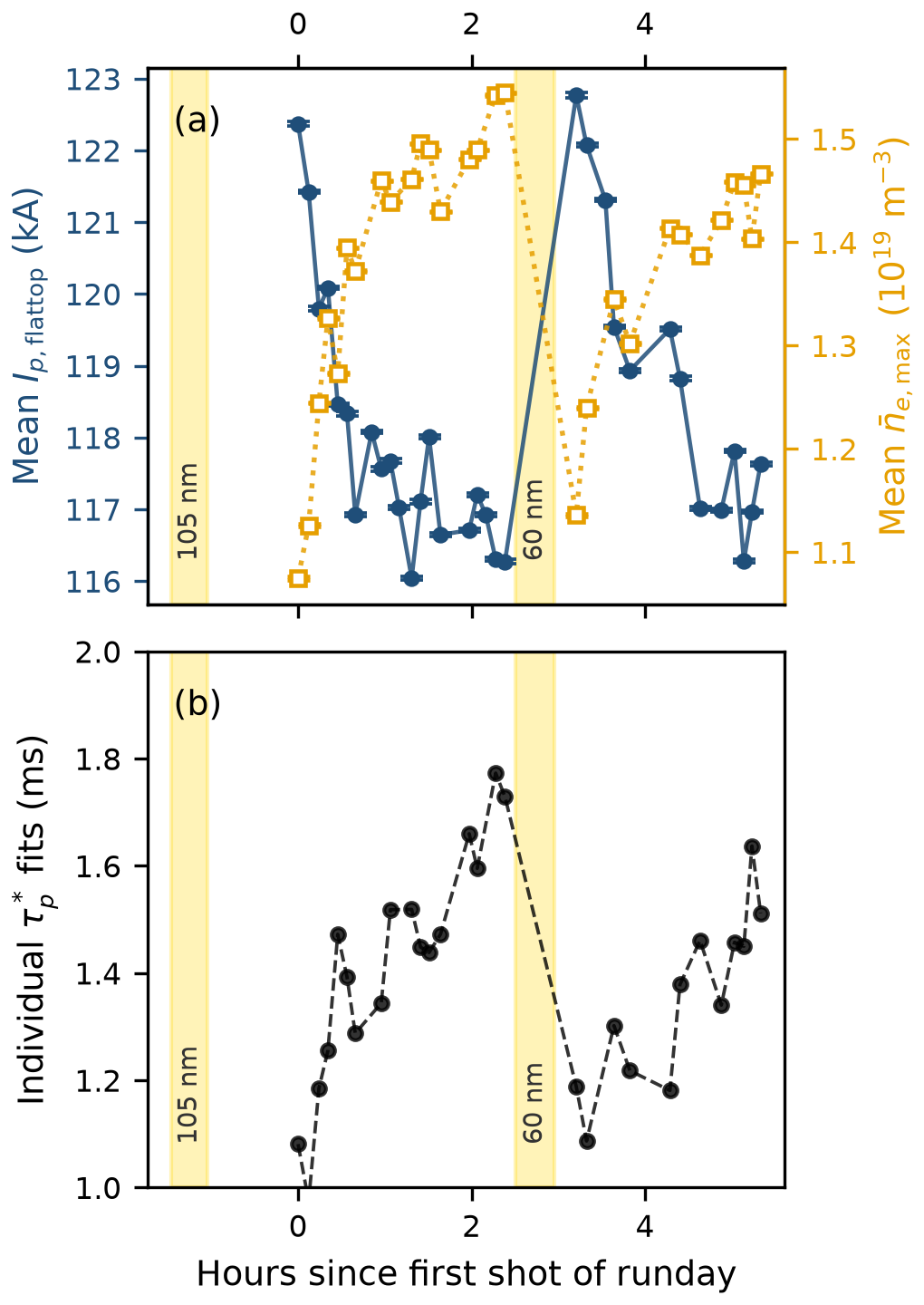}
\caption{\label{fig:sep22-operational-evolution} Evolution of a repeated low recycling discharge \cite{maan2024estimates} during a single run day. The run day was paused for a second evaporation midway through the run day with the Mark-II evaporator. (a) Evolution of mean flattop plasma current and peak line averaged density for the shot. Plasma current is shown by a solid line with filled circles on the left axis; line averaged density is shown by a dotted line with open squares on the right axis. (b) Individual \tps fits for shots after the evaporation at the beginning of the run day and after the second evaporation. The shaded bands indicate the two lithium evaporations.}
\end{figure}

Previously, comparisons of \ltx shots with identical field coil and gas programming on different run days showed that performance was significantly improved by fresher and larger lithium depositions \cite{maan2024estimates}. An \ltx run day typically comprised up to 60 repetitions of a single discharge to build an ensemble average, with plasma operation preceded by a lithium evaporation at the beginning of the run day. Figure~\ref{fig:sep22-operational-evolution} shows the evolution of plasma current and density for a high performance discharge with a documented ensemble averaged recycling coefficient of $R \sim 0.5$. Two evaporations were used on this run day. The first evaporation deposited an average coating thickness of 105~nm, followed by a second evaporation of \(\sim57\)~nm midway through the run day. 

Ensemble average plasma current, density, core density, and temperature for the low recycling discharge are presented in earlier work along with a comparison against the ensemble averages of the other two higher recycling cases \cite{maan2024estimates}. The time dependence of the plasma parameters and density decay time, \tps, fits during the run day was not discussed there. For \ltx, \(\tau_p^*\) denotes the effective particle confinement time inferred after external fueling is terminated: \(\tau_p^*\equiv \tau_p/(1-R)\simeq -N/(dN/dt)\), where \(N\) is the plasma particle inventory and \(R\) is the global recycling coefficient. The values presented here and in earlier work \cite{maan2024estimates}, are obtained from exponential fits to the line averaged density decay after the \ supersonic gas injector (SGI) fueling was turned off. Figure~\ref{fig:sep22-operational-evolution} shows that time dependence for the lowest recycling case's run day. Shots immediately following the first evaporation had higher plasma current and lower line averaged density than shots taken later in the day, before the second evaporation. Since \ltx ran without feedback control, the field coil programming and gas requests were identical for all shots. Higher plasma current at lower density is consistent with higher electron temperature. As more shots were fired, the plasma current flattop relaxed to a lower value and the line averaged density increased, indicating degraded plasma performance. The pattern repeated after the second evaporation. Shots immediately after the second evaporation again had higher plasma current and lower line averaged density than later shots. The \tps fits evolved with the plasma current and line averaged density: \tps was lower immediately after each evaporation and relaxed to higher values as more shots were fired. In the absence of fueling, the increase in \tps indicates weaker particle pumping by the lithium coated wall. This is consistent with the observed increase in line averaged density and decrease in plasma current. The plasma parameters and \tps fits therefore indicate that the wall condition evolved during the run day, and that performance degraded as the wall evolved away from the fresh lithium state created by evaporation. This interpretation is consistent with earlier \ltx observations that fresh and subsequent lithium evaporations increased plasma current and duration, reduced carbon and oxygen emission, and increased lithium emission. It is also consistent with molecular dynamics calculations showing that deuterium reflection and retention change with Li$_2$O layer thickness \cite{Maan_IEEE_2020,krstic2025oxide}.

\ltx operated in the lower range of high vacuum in the mid $10^{-8}$ Torr range. The dominant residual gas composition, measured with a residual gas analyzer (RGA), was hydrogen and either nitrogen or carbon monoxide, likely from a small leak \cite{Maan_thesis_2020}. Typically, the water RGA feature decreased after lithium evaporation and the hydrogen feature increased. This response is attributed to lithium on the PFCs cracking water into its constituents and retaining oxygen. It is consistent with laboratory observations of Li$_2$O formation on lithium when the total H$_2$O exposure is below 100 langmuirs (L). A langmuir is a vacuum exposure unit, 1 L = 1$\mu$Torr $\times$ 1 sec. RGA data indicate that water in the vacuum vessel was \textless $1 \times 10^{-9}$ Torr. The lithium coatings deposited by the evaporators therefore evolved in a residual water and hydrogen background, which drove lithium oxidation and surface lithium oxide formation. The plasma facing lithium evolved from fresh lithium toward an oxidized state during the run day \cite{Maan_IEEE_2020,Maan_PPCF_2020}. This surface evolution was diagnosed with in vacuo X-ray photoelectron spectroscopy (XPS) measurements of the lithium coated shells. Those measurements showed that lithium oxidation began within the first few hours after evaporation and that the lithium oxide layer grew over the first several hours until it exceeded the XPS probe depth \cite{Maan_PPCF_2020,Maan_IEEE_2020}.

Oxide growth on metals is a well studied process and can be modeled as advection and diffusion of metal and oxide atoms through the metal/oxide layer \cite{xu2012metal-oxidation}. XPS measurements can constrain the model estimated growth rate for \ltx at least up to the XPS probe depth, which is about 10 nm. The Li/O ratio in LTX XPS data evolves toward the 2:1 stoichiometry expected for Li$_2$O and saturates once the oxide thickness exceeds the XPS probe depth. For the \ltx after lithium water flux, \(5.6\times10^{15}\)~m\(^{-2}\)s\(^{-1}\), and unity sticking coefficient, one oxide monolayer forms in roughly 13~min. The XPS probe depth was saturated by about 3~nm of oxide after \(\sim5\)~h. Applying the thin film metal oxidation model to these data gives an initial Li$_2$O growth rate of 0.72~nm/h and a rate of 0.55~nm/h near the XPS probe depth \cite{xu2012metal-oxidation,Maan_thesis_2020}. A 100~nm coating would therefore take roughly 180~h to fully oxidize at the slower rate, but the chemically active near surface region changes within the first few hours.

\begin{figure}[htbp]
\centering
\includegraphics{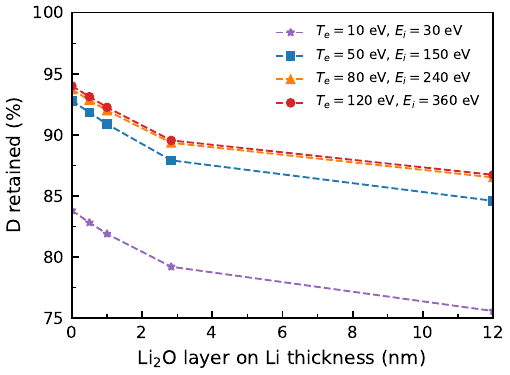}
\caption{\label{fig:krstic-retention}Estimated deuterium prompt retention as a function of Li$_2$O layer thickness on Li. Retention is calculated as one minus the prompt reflection probability \cite{krstic2025oxide}, with impact energy estimated as $E_i \simeq 3T_e$.}  
\end{figure}

Under residual water exposure relevant to LTX, Li$_2$O forms before the surface transitions toward LiOH. Longer water exposure drives the transition toward LiOH. An exposure of 100~L of H$_2$O corresponds to about 14~h under these conditions; therefore, evaporation at the beginning of the run day does not leave enough time for the lithium to evolve to LiOH\cite{Maan_IEEE_2020,Maan_thesis_2020}. To probe the dependence of recycling on thin Li$_2$O films that grow on top of elemental lithium a classical molecular dynamics study with ReaxFF Li-O-H potential was performed\cite{krstic2025oxide}. Amorphous lithium slabs were oxidized by adding O atoms one layer at a time, relaxed at 300~K, and then irradiated with normally incident 5--300~eV D atoms to calculate prompt reflection probabilities. Figure~\ref{fig:krstic-retention} converts those fitted reflection probabilities to retained fraction, \(100-P_R\), using \(E_i\simeq3T_e\).  Although H and D have similar chemical interactions with Li and Li$_2$O, their different masses produce kinematic isotope effects in reflection, implantation, and retention. Earlier ReaxFF calculations using the same interaction potential for H, D, and T, with only the isotope mass changed, found that prompt reflection from amorphous Li$_2$O decreases with isotope mass below approximately 25~eV, while the trend reverses at higher energies \cite{krstic2023detailed}. These results suggest that H may show a stronger prompt reflection response than D in the lower energy part of the incident distribution; however, quantitative H calculations for the evolving thin Li$_2$O on Li geometry in Fig.~\ref{fig:krstic-retention} remain future work. The prompt retention calculations in Fig.~\ref{fig:krstic-retention} connect this surface chemistry to the recycling response. Thin Li$_2$O layers on Li still retain a large fraction of incident deuterium, but the retained fraction changes as the oxide layer thickens \cite{krstic2025oxide}. The most favorable wall state is therefore likely accessed soon after evaporation, when the plasma facing surface is still lithium rich with only a thin oxide layer. Figure~\ref{fig:sep22-operational-evolution} shows that plasma current, density, and \tps changed over the first few shots after each evaporation. Accessing this state reproducibly requires lithium delivery on the same timescale as the \ltx shot cycle, which was about 5~min.  Although technically feasible, the evaporation midway through the run day was exceptional because it required an operational interruption of order 2~h and was therefore not compatible with routine between shot conditioning. This limitation motivates a remotely actuated evaporator that can refresh the coating on the normal shot cycle timescale.

Background gas chemistry and plasma driven sputtering modify the \ltx coating through distinct channels. Residual H$_2$O and oxygen-bearing gases convert elemental Li first to Li$_2$O and later LiOH, whereas plasma impact redistributes material during each discharge. Low energy calculations show that amorphous Li$_2$O sputters mainly Li, oxygen ejection is rare, so exposure can preferentially deplete Li and leave an oxygen-rich surface \cite{krstic2023detailed}. The oxide-layer molecular-dynamics data give a layer averaged yield near 100~eV of 0.065 target atoms per incident D and an upper envelope over 5--300~eV of 0.085 \cite{krstic2025oxide}. At an upper-bound \ltx flux of $10^{23}$~D~m$^{-2}$~s$^{-1}$, a 50~ms discharge gives a fluence of $5\times10^{21}$~D~m$^{-2}$. For a Li$_2$O atomic density of $1.22\times10^{29}$~m$^{-3}$, these yields imply 2.7--3.5~nm equivalent gross erosion, or 27--35\% of a 10~nm layer. Sputtered Li$^+$ can be driven promptly back to the surface by the sheath, partially offsetting net loss \cite{abe2023quantitative}. Because lithium has a near-unity sticking probability on cold solid surfaces, material not returned locally will tend to condense at the next surface it encounters \cite{farmer2009lithium-adsorption}, redistributing it away from high-flux limiter regions. This redistributed lithium may contribute to pumping elsewhere, but the net effect is likely reduced recycling control where it matters most.

This partial recycling replenishes the limiter coating less effectively than a new evaporation, which adds fresh lithium inventory. The 27--35\% value therefore bounds gross local removal under peak flux; NSTX-U awaits device-specific flux-energy distributions and redeposition modeling.

These observations establish the PMI requirements that organize the evaporator development described next. The source must refresh lithium on operational timescales, reproducibly coat the plasma wetted and limiting PFCs, and minimize oxygen, water, and related to handling impurity exposure. In this framing, low thermal mass, spatial coverage, controlled inventory, wetting, and in vacuo loading are methods for controlling the plasma-material interface rather than independent hardware optimizations.

\section{\label{sec:evolution}Evaporator design evolution on LTX and \ltx}

The design sequence below is organized by the requirements derived from PMI and established in Section~\ref{sec:shots}. Each evaporator iteration addresses a limitation in coating freshness, coverage, repeatability, impurity exposure, or molten lithium control that affects the condition presented to the plasma.

\begin{figure}[htbp]
\centering
\includegraphics[width=0.40\columnwidth]{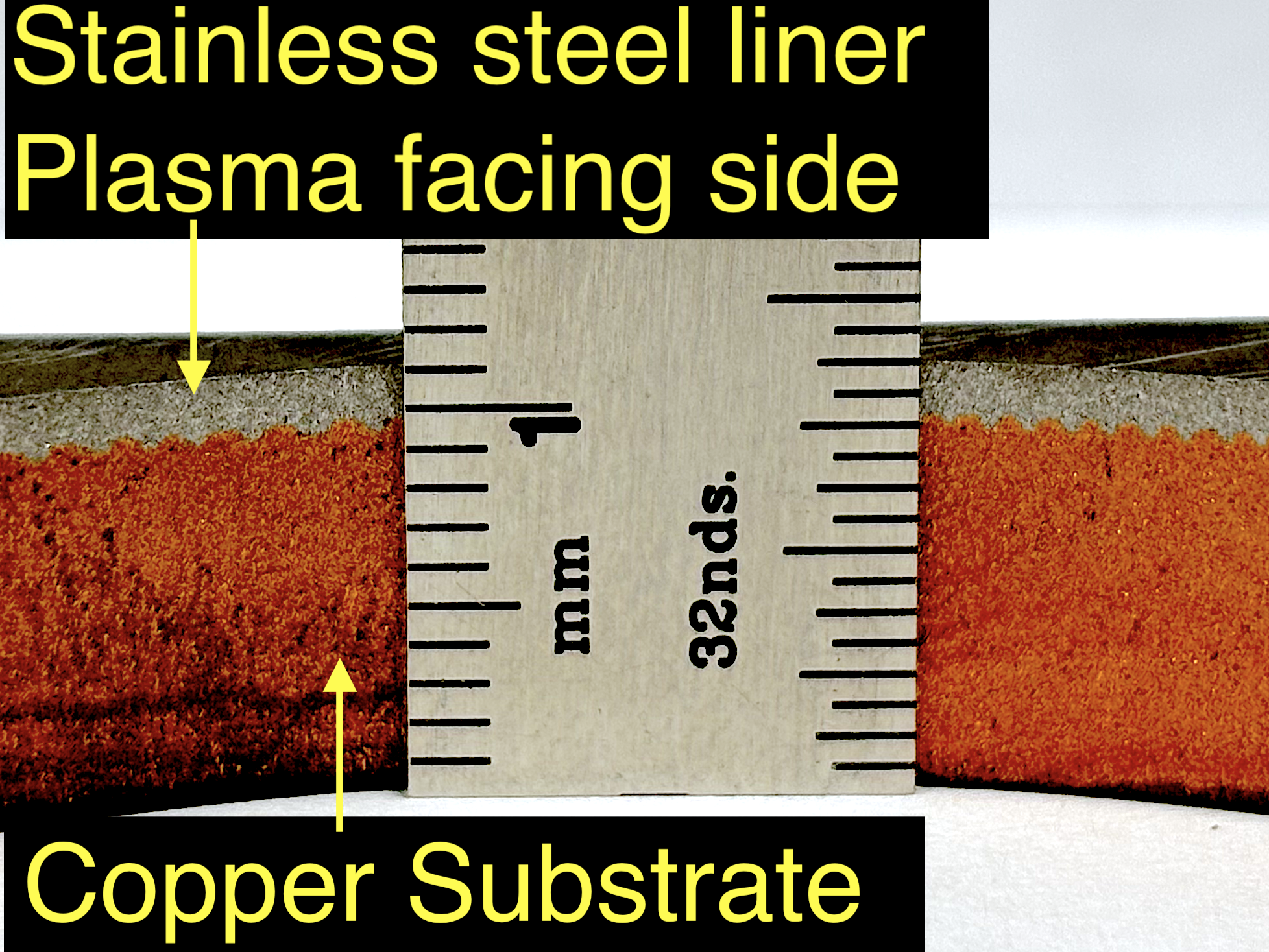}
\caption{\label{fig:shell} LTX and \ltx PFC shell material cross section. The top layer is the $\sim$ 1.5 mm thick stainless steel plasma facing surface, and the substrate is Oxygen-Free High Conductivity (OFHC) copper. The copper substrate is nickel plated to protect against accidental Li exposure.}
\end{figure}

Lithium delivery on LTX and \ltx evolved through several techniques before the low thermal mass evaporators that are the focus of this paper. In all cases, Li was applied to the stainless steel plasma facing side of the shell structure within the LTX/\ltx vacuum vessel. A representative cross section of the shell material is shown in Fig.~\ref{fig:shell}. The close fitting shell is divided into four quadrants: two upper and two lower shell sections. The shell shape was chosen to be conformal to the largest plasma that could fit within the vacuum vessel, with major radius $R=0.4$~m and minor radius $a=0.26$~m, so the lithium coated shell itself defined the plasma boundary \cite{Maan_IEEE_2020,maan2024estimates}.

The original LTX evaporators used yttria crucibles, loaded with solid lithium, that were inserted into the torus and resistively heated with tantalum heaters. Two evaporators were separated toroidally by 180 degrees. Typical evaporations used several grams of lithium, lasted 2--3 h, and were performed with a helium fill of about 5 mTorr to redistribute lithium vapor over the upper and lower shell surfaces. Some early tests used He glow discharges but mainly neutral He gas was used. DEGAS2 calculations and visual inspection indicated that coating uniformity depended on fill pressure \cite{stotler1994degas2,abrams}. Higher fill pressures shortened the collisional mean free path between lithium and helium atoms and produced more uniform coatings~\cite{schmitt-ltx-results-2013,majeski-ltx-particle-control-2013}. The method produced large area lithium coatings and was associated with improved LTX breakdown and plasma current relative to before lithium operation. However, it also required pumping out the helium and cooling the crucibles before plasma operation, adding a 1--2 h delay between evaporation and discharges~\cite{schmitt-ltx-results-2013,lucia-thesis-2015}. The helium assisted process also carried an impurity history penalty: the long evaporation, pumpout, and crucible cooling sequence left fresh lithium exposed to residual vacuum oxygen and water, and repeated use of the process introduced adsorbed water or oxygen bearing species during coating cycles \cite{lucia-thesis-2015}. The He gas itself also introduced impurities from both the feed lines and the gas supply; even ultrahigh purity He containing parts per million impurities introduced more residual gas than the base vacuum background. 

These LTX results also provide the rationale for not adopting plasma-assisted lithium conditioning in the present design. Early trials included lithium evaporation during He glow discharge cleaning (He-GDC), although most helium-dispersed coatings were produced with neutral He. The helium-dispersed coating lineage did not translate successfully to liquid operation, when the coatings were liquefied, absorbed impurities, especially oxygen, were released, recycling increased, and plasma performance degraded. Rapid electron-beam evaporation, which was used on LTX in later campaigns, did produce cleaner liquid surfaces and energy confinement up to an order of magnitude above the helium-dispersed case \cite{majeski-ltx-particle-control-2013,schmitt-ltx-liquid-walls-2015}. Pure thermal evaporation without a concurrent He discharge was therefore preferred for \ltx and f-LITER to minimize process-gas exposure and preserve a fresh liquid-lithium surface in \ltx experiments.

\begin{figure}[htbp]
\centering
\includegraphics[width=\columnwidth,height=0.58\textheight,keepaspectratio]{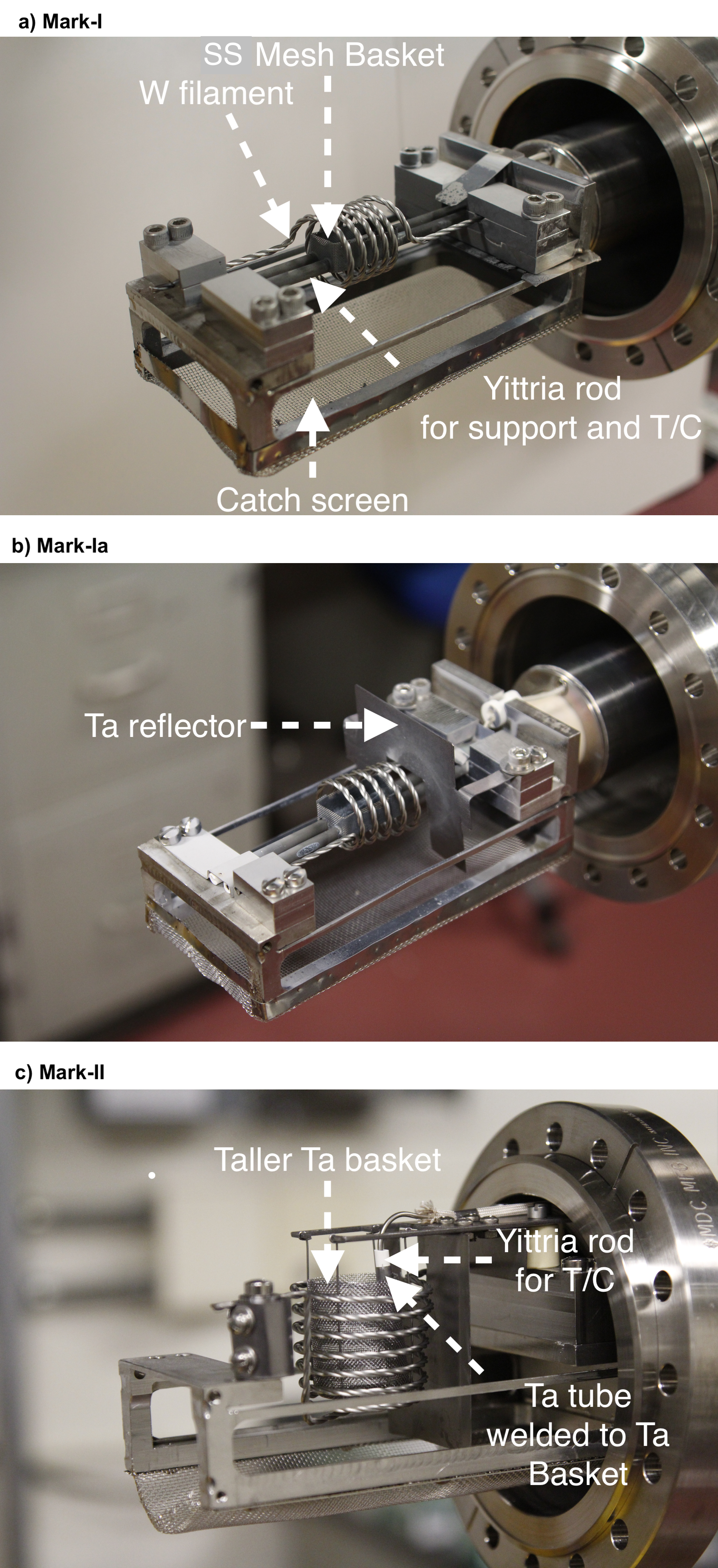}
\caption{\label{fig:mark-evaporators}Photographs of the evaporator design progression: (a) Mark-I, (b) Mark-Ia, and (c) Mark-II.}
\end{figure}

In later LTX operation, lithium was introduced with a liquid filler system, forming small liquid pools in the lower shells that were then evaporated by electron beams steered with the LTX magnetic field coils \cite{lucia-thesis-2015}. This approach removed the helium backfill step, but it introduced a different constraint of requiring preheating of the lower shells. The LTX plasma facing shells were stainless steel on the plasma facing side, explosively bonded to copper (see Fig. \ref{fig:shell}). This construction was chosen for lithium compatibility at the surface and high thermal and electrical conductivity in the shell to aid in passive magnetohydrodynamic stability against large scale global toroidal modes \cite{garofalo1999-external-kink-rwm,fitzpatrick1996-fake-rotating-shell,berzak-hopkins2012-ltx-equilibrium}. Lithium in the lower shell reservoirs was found to be strongly heat sunk into the shell structure \cite{lucia-thesis-2015}. Evaporating lithium from the shell therefore required preheating the lower shells to about 300~$^\circ$C before filling and electron beam operation. Because of the high thermal mass of the shells, heating and cooling took hours, and hot shell operation kept the base vacuum elevated for long periods, leading to an impurity codeposition problem \cite{schmitt-ltx-results-2013,lucia-thesis-2015}. The electron beam system reduced the helium process delay and process gas impurities, but deposition remained sensitive to residual vacuum exposure and surface oxidation. In situ surface analysis on LTX showed that these evaporated coatings evolved rapidly in the residual vacuum and that plasma performance depended on surface condition as well as the amount of lithium introduced \cite{lucia-ltx-surface-2015}. These constraints motivated the subsequent move on \ltx toward low thermal mass flash evaporators designed for faster, more flexible, and more repeatable lithium application, with the eventual goal of between shot operation.

\ltx went through three iterations of the flash evaporator design. Flash evaporation largely solved the impurity codeposition issue \cite{Maan_IEEE_2020} because the evaporation cycle was over in a few minutes. Subsequent design iterations were driven by the need to improve coverage, retain molten lithium in the basket without dripping, and reduce lithium oxidation before and during transfer of lithium to the evaporator. Table~\ref{tab:evaporator-evolution} summarizes the design changes, advances, and remaining limitations. Photographs of the three \ltx flash evaporator generations are shown in Fig.~\ref{fig:mark-evaporators}.

\begin{table*}[htbp]
\centering
\caption{LTX flash-evaporator design evolution.}
\label{tab:evaporator-evolution}
\small
\setlength{\tabcolsep}{4pt}
\renewcommand{\arraystretch}{1.18}
\newcommand{\evolcell}[2]{\parbox[t]{#1}{#2}}
\begin{tabular}{llll}
\hline
\evolcell{0.14\textwidth}{\textbf{Generation}} & \evolcell{0.25\textwidth}{\textbf{Main change}} & \evolcell{0.27\textwidth}{\textbf{Problem addressed}} & \evolcell{0.25\textwidth}{\textbf{Remaining limitation}} \\
\hline
\evolcell{0.14\textwidth}{Mark-I} & \evolcell{0.25\textwidth}{Low-mass mesh basket and W heater} & \evolcell{0.27\textwidth}{Slow evaporation and impurity codeposition} & \evolcell{0.25\textwidth}{High-field-side shadowing; manual loading} \\
\evolcell{0.14\textwidth}{Mark-Ia} & \evolcell{0.25\textwidth}{Less-obstructive carriage and Ta reflector} & \evolcell{0.27\textwidth}{High-field-side coverage} & \evolcell{0.25\textwidth}{Foil loading, venting, and long pumpdown} \\
\evolcell{0.14\textwidth}{Mark-II} & \evolcell{0.25\textwidth}{Larger Ta basket, lower aspect ratio, and fiber-felt liner} & \evolcell{0.27\textwidth}{Inventory, coverage, and dripping} & \evolcell{0.25\textwidth}{Loading still interrupted operation} \\
\evolcell{0.14\textwidth}{Dropper-fed Mark-II} & \evolcell{0.25\textwidth}{In-vacuum liquid-Li feed; controlled 2--3~g loads} & \evolcell{0.27\textwidth}{Handling exposure and inventory control} & \evolcell{0.25\textwidth}{NSTX-U scale, rate, and lifetime unqualified} \\
\hline
\end{tabular}
\end{table*}

\begin{figure}[htbp]
\centering
\includegraphics[width=\columnwidth]{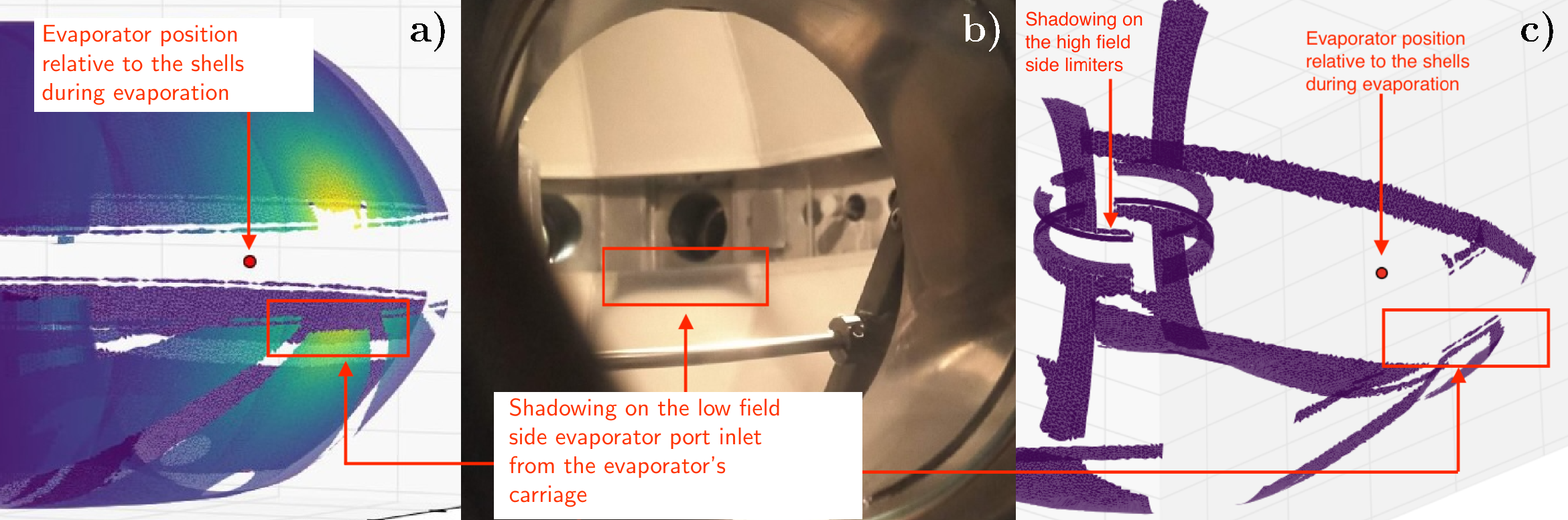}
\caption{\label{fig:mark-i-coverage}Mark-I lithium coverage and coverage limitation: (a) predicted line of sight coating pattern, (b) lithium deposition pattern as observed on the low field side North shell, and (c) predicted line of sight occlusion from the evaporator structure.}
\end{figure}

The first \ltx low thermal mass evaporator, here referred to as Mark-I and shown in Fig.~\ref{fig:mark-evaporators}(a), was the first flash evaporator installed on \ltx. It replaced long lead time shell heating with a compact source that could be inserted near the poloidal center of the vessel. The evaporator used a stainless steel mesh basket supported by yttria rods, with a tungsten coil heater surrounding the basket and a Type K thermocouple inserted through one of the supports to monitor temperature \cite{Maan_IEEE_2020,elliott2020initial}. The basket was made from three pieces of stainless steel mesh that were bent to shape and spot welded. The mesh had approximately 64 openings per inch and a wire diameter of about 0.2 mm, corresponding to a nominal pitch of 0.40 mm and an approximate clear opening of 0.20 mm. Solid lithium foil pieces, 1.5 mm thick, were cut from a spool of lithium foil and manually loaded into the evaporator, after which the assembly was pumped down and inserted into \ltx. Two evaporators were installed at diametrically opposite toroidal locations, and each was positioned under a shell penetration with line of sight to a quartz crystal microbalance (QCM) used to monitor deposited lithium during evaporation \cite{Maan_IEEE_2020}. This design reduced the heated mass from both bottom shells to the evaporator head and provided a practical path for faster, repeated lithium application during run preparation. 

During evaporation from Mark-I evaporators, the \ltx base pressure rose from the low \(10^{-8}\)~Torr base vacuum to, at most, the low \(10^{-6}\)~Torr range. Even at the upper end of these pressures, the lithium mean free path is much larger than the \ltx vacuum vessel dimensions. Lithium vapor incident on a room temperature surface is also expected to have a high, near unity sticking probability \cite{farmer2009lithium-adsorption}. Deposition from the evaporator can therefore be treated as a molecular flow, line of sight coating process, without significant scattering or reemission from surfaces.  

Because the basket dimensions were much smaller than the vessel, lithium deposition on the shells was modeled with a point source evaporation approximation. Lithium is treated as originating from an effectively infinitesimal source. Deposition on the shell surface then scales with \(\cos\theta\), and inversely with the square of the source to surface distance \cite{ohring2002thin-films}. Because lithium has a near unity sticking probability on cold metal surfaces, surfaces that cannot see the source receive no direct coating. The evaporation model constrained the source strength with the QCM deposited thickness. The model then estimated deposition on each discretized shell element from the source to element distance and incidence angle, only where a direct line of sight existed. A ray tracing calculation added occlusion from evaporator structural elements, the LTX shell geometry, and the center column. The model was therefore used to predict the spatial distribution of lithium on the shells and to identify where the Mark-I evaporators could not reach, as shown in Fig.~\ref{fig:mark-i-coverage}(a). Comparing predicted deposition patterns with after evaporation photographs showed reasonable agreement between predicted and observed coverage, Fig.~\ref{fig:mark-i-coverage}(b). 

This point source model is an engineering estimate rather than a quantitative full wall validation. The long lithium mean free path and near unity sticking probability support the line of sight treatment, and the model reproduces the observed shadowing pattern in Fig.~\ref{fig:mark-i-coverage}(b); however, the angular emission distribution and absolute local deposition remain to be validated. Following completion of tokamak operations, the \ltx vacuum vessel will be used to test the NSTX-U evaporator deposition footprint and engineering interlocks. Comparison with in vessel deposition measurements, including diagnostic window exposure, will be reported separately.

The model predicted that roughly 86\% of the shell surface area was coated by the Mark-I evaporators. This estimate was probably conservative based on after evaporation photographs, but it identified where coverage was missing \cite{Maan_thesis_2020}. Because evaporation from the compact basket was essentially line of sight, structural elements in the evaporator carriage blocked trajectories to parts of the plasma facing shells, as shown in Fig.~\ref{fig:mark-i-coverage}(c). This mattered because the high field side shell region is where \ltx plasmas typically limited. The evaporators therefore failed to coat the most operationally important surface. 

Mark-Ia evaporators were designed to overcome this coverage limitation, as shown in Fig.~\ref{fig:mark-evaporators}(b). Some of the front structural members that had shadowed the high field side shell were removed, and a tantalum reflector was added to redirect lithium vapor toward the high field side limiting surfaces. Tantalum was selected because pure tantalum has excellent resistance to dissolution attack by liquid lithium at temperatures well above the Mark-Ia operating range. There was no significant risk of oxygen driven lithium penetration of tantalum leading to dissolution \cite{distefano1964refractory-lithium} because \ltx operated in the lower range of high vacuum (~$5 \times 10^{-8}$ Torr), and with lithium coatings, oxygen exposure was generally limited to ~ 10$^{-9}$ Torr or less \cite{maan2024estimates, boyle_nme}. The reflector could also be fabricated as a thin formed part because tantalum is a ductile, easily fabricated refractory metal \cite{schulz2014niobium-tantalum}. The resulting Mark-Ia evaporator was essentially a Mark-I evaporator head with most of the shadowing elements removed and a lithium compatible tantalum reflector added to improve coverage of the high field side limiters \cite{MAAN2023_NME}.

Both Mark-I and Mark-Ia evaporators were loaded with 1.5 mm thick solid lithium foil. The loading process allowed charges up to 2~g to be moved from the argon glovebox to the LTX-\(\beta\) test cell. The lithium foil was cut into rectangular coupons that were slightly smaller than the basket and stacked on top of each other. The foil surfaces were scraped to remove accumulated oxides, which helped the coupons stick to each other. Once prepared and weighed in the argon glovebox, the Li charge was sealed in an argon filled transfer container and carried to the \ltx test cell. With the lithium evaporator assembly attached to \ltx and retracted behind a gate valve, the assembly was backfilled with argon, and a flange directly above the evaporator head was removed. The lithium charge was taken from the transfer container with tweezers or a similar tool and inserted into the woven wire basket while maintaining a flow of argon through the assembly, in from the argon bottle, out from the access port. The flange was resealed, the evaporator assembly was pumped down, and the gate valve to \ltx was opened only after a suitable base pressure was reached. The entire process took at least 8 hours, or an entire work day. This process reduced air and moisture exposure compared with open handling, but it was ultimately limited by the need to handle and position the solid lithium pieces, carefully and quickly. Anecdotally, the duration of air exposure during loading was observed to affect the pumpdown time after loading.

\begin{figure}[htbp]
\centering
\includegraphics[width=\columnwidth]{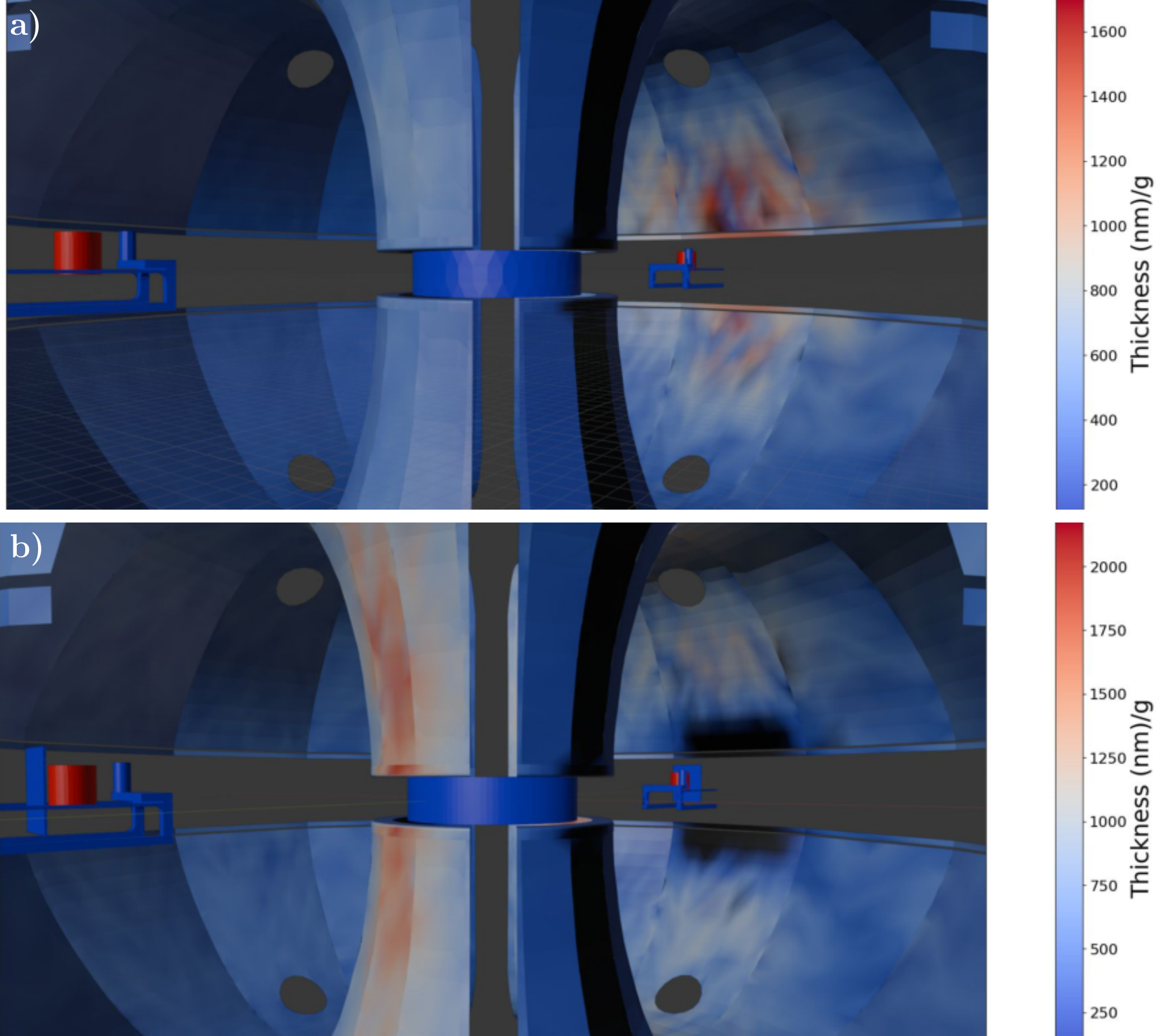}
\caption{\label{fig:mark-iii-coverage} Predicted Mark-II lithium coating thickness: (a) without the tantalum reflector and (b) with the tantalum reflector. A diverging color scale is used because it distinguishes regions with no or negligible deposition from regions with low deposition while preserving the enhanced deposition on the high field side surfaces. Each panel uses the thickness scale shown at right.}
\end{figure}

The Mark-II basket replaced the Mark-I/Ia stainless steel woven mesh basket with a taller, wider, and coarser tantalum mesh basket and reoriented the heater vertically to allow easier access to the basket from above. The tantalum mesh had approximately 38 openings per inch and a wire diameter of about 0.12~mm, corresponding to a nominal pitch of 0.67~mm and an approximate clear opening of 0.55~mm. The larger basket was intended to increase the usable lithium inventory, improve line of sight reach to the remaining undercoated shell regions, Fig.~\ref{fig:mark-iii-coverage}, and reduce the need for frequent vents to load the evaporator. The coarser mesh, however, allowed lithium to drip. Lithium droplets were observed on the mesh screen below the evaporator subassembly and on the port opening into the vessel. To avoid dripping, the Mark-II basket was lined with 316L stainless steel sintered metal fiber felt, which provided a wettable internal surface intended to hold molten lithium inside the basket during heating and evaporation. The felt is a nonwoven network of micron scale stainless steel fibers consolidated by high temperature vacuum sintering, producing metallurgical necks at fiber to fiber contact points rather than using an adhesive, braze, or polymer binder. Sintered 316L stainless steel fiber felts are therefore fully metallic, high porosity structures whose mechanical integrity is set by the sintering joints between fibers and the fiber ligaments themselves \cite{ma2018sintered-fiber-felt}. The felt liner was fabricated by cutting a piece of the felt to fit inside the basket. The felt lined Mark-II basket was then loaded with chunks cut from 0.5~in. diameter lithium rods in the same way as the coupons loaded into the Mark-I and Mark-Ia baskets. The felt liner prevented dripping during melting and evaporation. Inspection after evaporation showed that lithium had wetted into the felt structure, indicating that the felt provided a path to hold molten lithium in place during evaporation \cite{MAAN2023_NME}.

Despite the improved basket geometry and felt metal liner, Mark-II remained a manually loaded solid lithium evaporator. Lithium pieces still had to be prepared, transferred under inert gas, placed in the basket, and pumped down before operation, so exposure of lithium during transfer between the argon filled container and the argon vented evaporator still introduced impurities. The larger Li rods were more pure than the Li foils, and their lower ratio of surface area to volume reduced impurity accumulation during storage, handling, and loading. The taller basket increased the charge capacity and therefore reduced how often the evaporator assembly had to be vented and reloaded, but it did not remove the operational interruption associated with solid lithium handling. Mark-II evaporators represent the final solid lithium flash evaporator design in this development path: it solved the main geometric coverage and dripping limitations of the earlier sources, while leaving lithium loading and coating freshness as the remaining motivations for an evaporator fed by a dropper. 

\section{\label{sec:mark2}NSTX-U requirements and f-LITER design demonstration in \ltx}
\subsection{\label{sec:liter}Limitations of legacy NSTX downward LITERs}
The legacy NSTX lithium coating system used downward facing LiThium EvaporaRs (LITERs) aimed at the lower divertor. Two evaporators were located at toroidal angles of 165$^\circ$ and 315$^\circ$. Their central axes were aimed at the lower divertor, and the measured lithium angular distributions were used to estimate the coating footprint \cite{kugel2009evaporated-coatings}. This configuration was well matched to lower biased operation and produced useful plasma conditioning. Lithium coatings from the LITERs reduced plasma density, suppressed edge localized modes (ELMs), improved confinement, increased pedestal temperature, and reduced edge neutral density in NSTX discharges \cite{kugel2009evaporated-coatings,kugel2010lithium-coatings-nstx,Maingi_2012}. This geometry, however, did not provide the upper divertor coverage needed for NSTX-U balanced double null or upward biased operation \cite{roquemore2013upward-facing}.

One operational limitation of the LITERs was their thermal inertia. At evaporation temperatures above 450$^\circ$C, the LITERs took $\sim$ 2.5 h to cool down below 250$^\circ$C when the heaters were fully turned off. At 250$^\circ$C, the lithium evaporation rate becomes negligible. Because the NSTX discharge cycle was $\sim$ 20 min, the LITER sources were kept at evaporation temperature continuously during a run day at a selected evaporation rate. About 60~s before each discharge, they were retracted with a bellows drive behind lithium blocking shutters. The shutters stayed closed during the helium glow discharge conditioning (HeGDC) cycle to avoid lithium helium codeposition. They also protected diagnostic windows from lithium coating during discharges. After the discharge shutters opened, the LITERs were reinserted and lithium was deposited for 10~min on the lower divertor before the next shot \cite{kugel2009evaporated-coatings}. These operational choices meant that evaporated lithium was intercepted by shutters whenever the LITERs were retracted. Because the shutters were exposed to lithium buildup, they were a reliability risk. If a lithium coated shutter or its drive failed, repair required access to an in vessel component rather than a simple operating reset. NSTX documentation records refurbishment of a failed LITER shutter mechanism using an argon vent\cite{nstx2010-year-end-report}. Such a vent would interrupt NSTX-U operation and force recovery of the vessel vacuum conditioning before routine plasma operation could resume 

The shutter loss was also a lithium inventory penalty. A simple duty cycle estimate illustrates the scale, although historical NSTX shutter inventory was not directly measured. In a 9~h run day with a 20~min shot cycle, there are about 27 possible shots. If the hot LITER is inserted for 10~min between shots and parked behind the shutter for the remaining 10~min, about half of the evaporated lithium is intercepted by the shutter. For an 80~g reservoir that lasts two weeks, this corresponds to order 40~g of lithium deposited on the shutter rather than on plasma facing surfaces.

Once the LITERs ran empty, they were retracted from NSTX-U, isolated from the vessel, and vented. The probes were then lifted off the vessel with a crane and moved to a staging area. The LITERs were mounted on a dedicated fill station. Before refilling, the residual lithium was evaporated out of the reservoir and weighed. The LITER was heated to 220$^\circ$C while liquid lithium was transferred from a 320$^\circ$C liquid lithium canister. After filling, the reservoir and snout were heated to 600$^\circ$C for 5~min to outgas trapped argon and other gases from the lithium refill process. NSTX therefore required four LITER probes, two dedicated fill stands, and a crane to handle the probes during refilling. Once refilled, the probes were lifted back to the vessel, bolted, pumped down, and reinserted. The whole process took at least a full working day and typically extended overnight, since the reinstalled probe required a 12--16 h bakeout after the lift, refill, outgas, and reinstall sequence. The process was reliable, but it was time consuming and operationally complex, and had to be performed twice as often as it would if the shutter didn't collect half of the lithium. Limited PFC coverage, complex handling, and the risk of shutter failure motivated a new evaporator system for NSTX-U with broader coverage, faster between shot use, and more reliable operation.

\subsection{\label{sec:liq}Demonstration of f-LITER liquid lithium refill and operation in \ltx}
The new design must provide line of sight coverage to upper divertor, midplane, and center column PFCs for double null operation; reduce lithium loss to shutters or parked structures; preserve a fresh lithium surface on the timescale of the shot cycle; minimize oxygen, water, and argon contamination introduced during loading; hold molten lithium in the evaporator without dripping; and allow the evaporator head or lithium reservoir to be serviced without a major vessel intervention. These requirements define the f-LITER architecture described below.

\begin{figure}[htbp]
\centering
\includegraphics[width=\columnwidth]{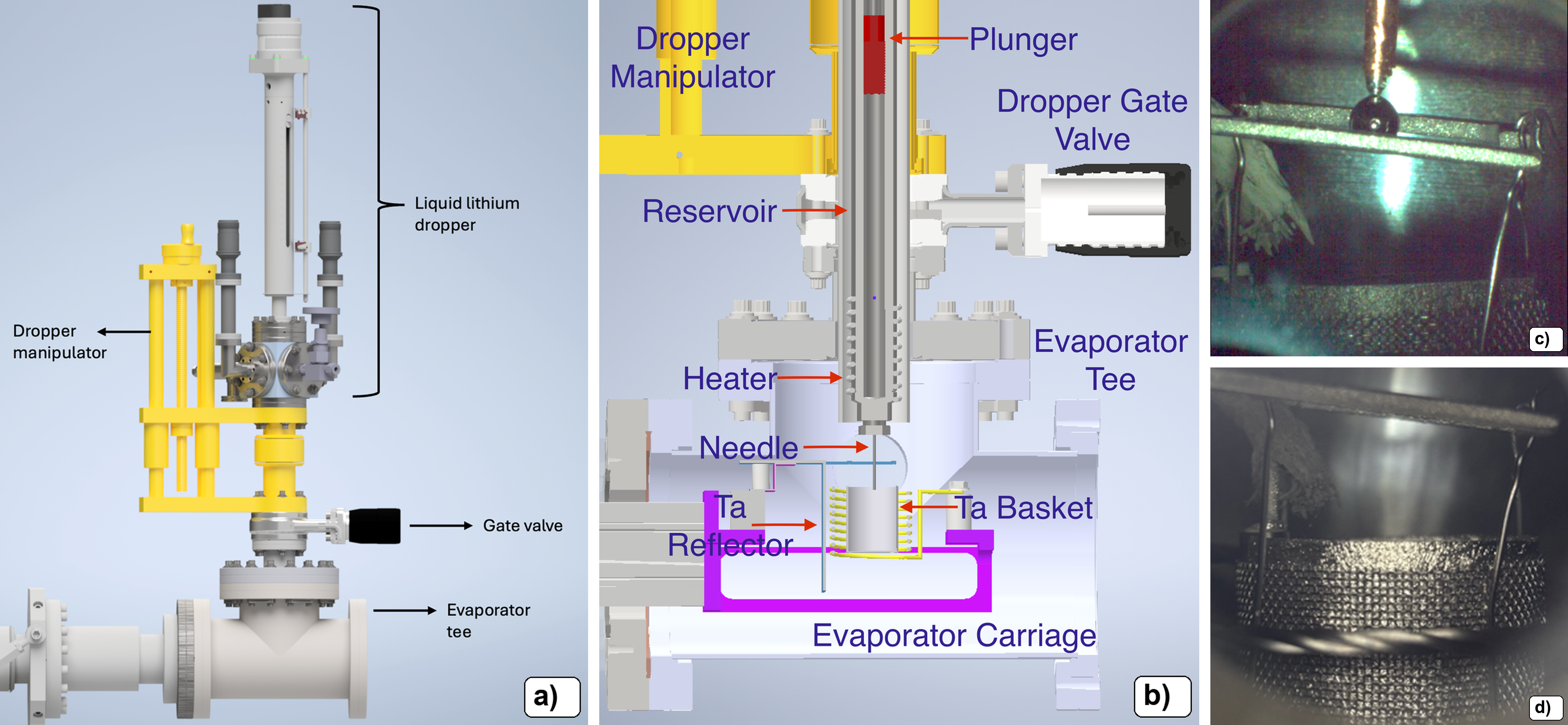}
\caption{\label{fig:mark-iv-dropper}Liquid lithium dropper and Mark-II evaporator assembly tested on \ltx and in DART. (a) Full dropper assembly showing the dropper manipulator, liquid lithium dropper, and evaporator tee. (b) Cross section view with key dropper and evaporator head components labeled. (c) Photograph of a liquid lithium droplet suspended from the dropper needle above the basket during loading in DART, before the basket is wet. (d) Photograph of the basket after wetting and evaporation in DART; compared with panel (c), the felt metal liner has a shiny metallic appearance, indicating complete wetting. Panels (c) and (d) document the proof of loading and proof of wetting tests in DART.}
\end{figure}

The remaining limitation of the \ltx evaporator lineage was lithium loading. Solid lithium handling was cumbersome on both LTX and \ltx. The Mark-II basket, felt metal liner, and low thermal mass head had already addressed the earlier coverage and dripping problems. The next step was to replace manual solid lithium loading with an in vacuo liquid lithium dropper based on a recently developed liquid lithium dropper design \cite{lopezperez2023li-dropper}. The dropper is small enough to be easily reloaded in a glovebox and so avoids transferring solid lithium in air. Loading solid Li and then operating in vacuum also avoids filling hot liquid lithium into the source under argon flow, as was done for the NSTX LITERs. 

A Mark-II evaporator modified to accept such a dropper was tested on \ltx and on the Deposition Analysis and Reliability Testbed (DART), a dedicated vacuum chamber designed for testing evaporators. The evaporator and dropper combination was mounted on DART to test the lithium loading and evaporation sequence, Fig.~\ref{fig:mark-iv-dropper}(c,d). The dropper reservoir is a smooth bore 316 stainless steel tube. It is 13 inches long, with 0.635 inches inner diameter and 0.875 inches outer diameter. The top one inch is threaded 7/8-14 Unified National Fine (UNF) and mates to a custom double sided 2.75 inch ConFlat (CF) flange on a 2.75 inch CF cube. A 2.75~in. to 1.33~in. CF reducer above the flange carries a UHV-Design linear bellows drive. The drive shaft couples to a 316 stainless steel plunger with 0.628~in. outer diameter. A nozzle welded to the lower end of the body terminates in a 1/4~in. National Pipe Thread (NPT) connection for a stainless steel needle. The cube also carries thermocouple and power feedthroughs, a 1.33~in. right angle valve for pumping, and a needle valve for argon venting.

The cube assembly is mounted on a Thermionics bellows manipulator with 6~in. of vertical stroke. A protective tube runs along the body and keeps the heater and thermocouple leads from catching in the bellows during motion. The dropper sits above a 2.75~in. gate valve, so it can be pumped and vented independently from the evaporator. The dropper holds about 10~g of lithium and is loaded in a dedicated glovebox with O$_2$ and H$_2$O below 0.1~ppm. After transfer, the assembly is pumped to UHV. Heating is provided by a Heat Wave Labs coiled heater and controlled with a Eurotherm 3508 Advanced Temperature Controller and a TDK Lambda programmable power supply. A typical loading ramp is 5~$^\circ$C/s. During loading, the evaporator basket is held at 250$^\circ$C to warm the needle and promote wetting on the felt metal liner. Once the dropper body and needle reach 200$^\circ$C, the dropper is lowered so that the needle tip sits about 2~cm above the basket. The plunger is then actuated to dispense droplets into the basket. The DART and \ltx tests used 2--3~g loads per dropper cycle. At this inventory, most of the lithium wicked into the felt metal liner rather than forming a free pool in the basket, further reducing the chance of lithium dripping.

\enlargethispage{1.1\baselineskip}
Including testing in DART and operations in \ltx, the dropper was used continuously for seven months to load the lithium evaporator. Typical lithium drop sizes were ~1 cm in diameter. In total, the dropper was used to deposit ~25 g of lithium.

Neither lithium wicking along the needle exterior nor progressive nozzle clogging was observed during this testing. These are dropper failure modes; the dropper is isolated behind a gate valve and can be removed for service without venting the NSTX-U vessel or interrupting plasma operations.

\begin{figure}[htbp]
\centering
\includegraphics[width=0.45\columnwidth]{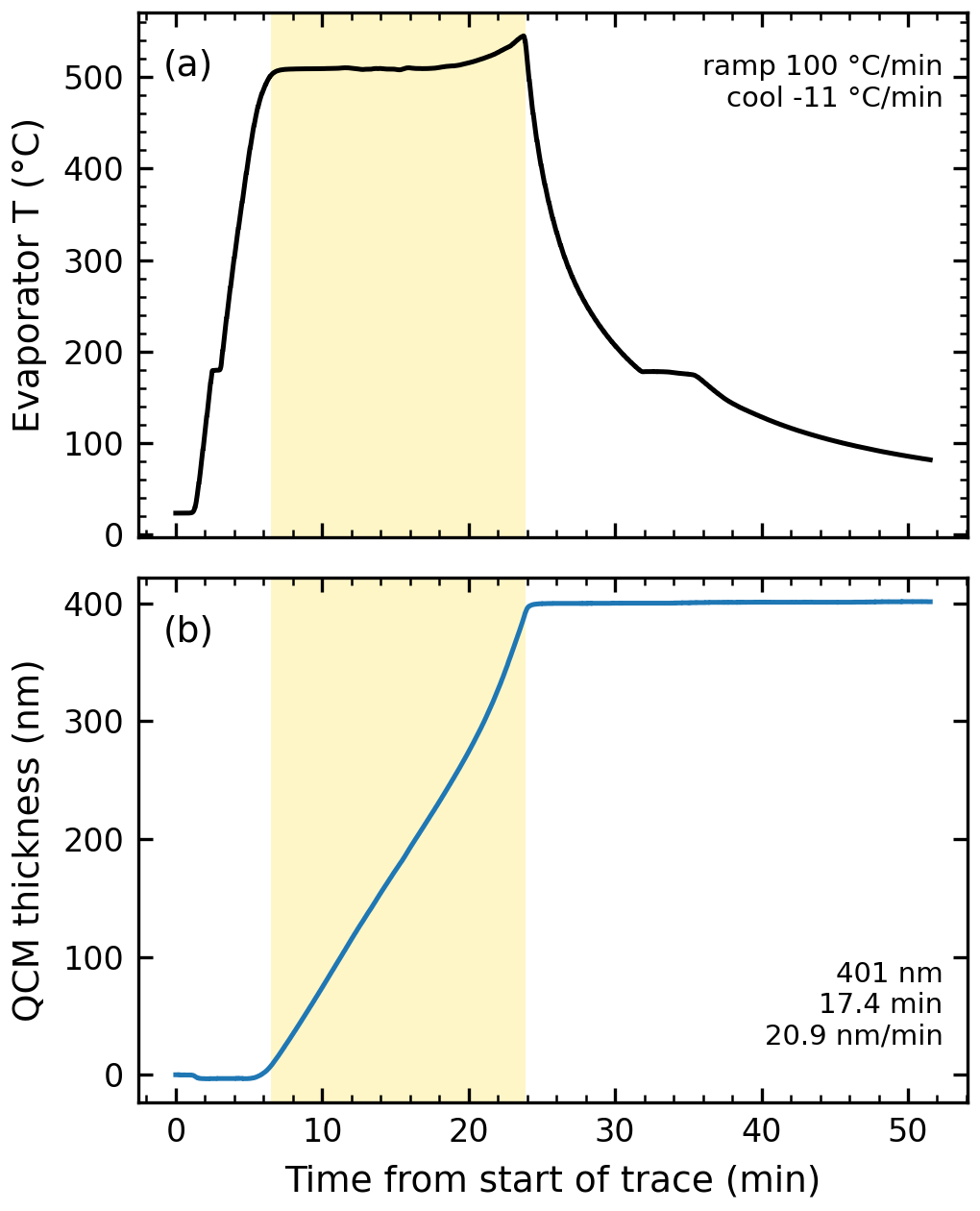}
\caption{\label{fig:dropper-qcm-evaporation}Representative \ltx evaporation at the beginning of the run day from a dropper loaded evaporator, using the evaporator temperature and QCM thickness from Dec. 11, 2025. (a) Evaporator temperature during the heating and cooling cycle. (b) QCM measured lithium thickness. The shaded region indicates the active deposition interval used to estimate the evaporation rate.}
\end{figure}

Figure~\ref{fig:dropper-qcm-evaporation} shows a representative evaporation at the beginning of the run day from the dropper loaded evaporator on \ltx. The evaporator reached a peak temperature of 545$^\circ$C. The temperature ramp rate between 100$^\circ$C and 450$^\circ$C was about 100$^\circ$C/min, and the cooldown rate through the same range was about 11$^\circ$C/min. The evaporator cooled from the end of active deposition to 250$^\circ$C in about 4.3~min. This is a practical shutoff temperature for between shot operation. Although lithium has a finite vapor pressure at 250$^\circ$C, the expected evaporation rate is negligible on the coating timescale compared with operation near 545$^\circ$C \cite{hicks1963vapor-pressure}. The QCM thickness increased by 401~nm over 17.4~min, corresponding to a shell averaged evaporation rate of 20.9~nm/min during the active deposition interval. At this rate, a 100~nm evaporation corresponds to about 4.8~min of active deposition, with longer cycles required when the evaporator contains a larger lithium inventory.

The thermocouple does not contact the liquid lithium directly; it is housed in a blind yttria tube held within a tantalum tube spot welded to the basket. It therefore responds through the basket and tube assembly. After heater shutoff, the small lithium inventory likely cools faster than this indirect sensor path, so the measured temperature lags the lithium temperature. The lithium vapor pressure also falls steeply, from about $1.2\times10^{-2}$~Torr at 545$^\circ$C to $2.8\times10^{-8}$~Torr at 250$^\circ$C, a decrease of more than five orders of magnitude \cite{hicks1963vapor-pressure}. The evaporative flux therefore rapidly falls below the measurable deposition rate during cooldown even though liquid lithium remains in the basket.

The \ltx installation contained two QCMs. A thermocouple tack welded to the holder of one QCM provided an independent measure of thermal loading on the crystal head, although it did not directly measure the quartz surface. The measurement was deemed necessary since QCM oscillating frequency can drfit due to thermal load and appear as spurious negative thickness. Figure~\ref{fig:dropper-qcm-evaporation} uses the thickness recorded by the other QCM. During the Dec. 11 evaporation, the thermocouple instrumented QCM holder showed temperature excursion from a pre-evaporation median of 21.95$^\circ$C to a maximum of 22.79$^\circ$C and returned to approximately 21.86$^\circ$C after evaporation, giving a measured excursion of only 0.84$^\circ$C. Crystal heating produces reversible frequency drift; source warmup usually causes the indicated thickness to decrease, while cooling produces an increase, and an unstable crystal may drift by as much as 10~Hz/$^\circ$C \cite{inficon_sqm160}. The simultaneous frequency and thickness records give a conversion of 4.61~Hz/nm, so this conservative specification bounds the corresponding apparent thickness shift at less than 1.8~nm, or less than 0.5\% of the 401~nm deposition. No explicit temperature correction was therefore applied. The total thickness was referenced to the pre-evaporation baseline and evaluated after the instrumented QCM-holder temperature returned close to baseline.

The same point source scaling used for the \ltx coverage estimates provides an order of magnitude NSTX-U cycle time estimate. The \ltx QCM was located about 26~in. from the evaporator, while the NSTX-U evaporator will be about 70~in. from the upper divertor. With inverse square scaling, and taking the incidence factor to be unity, the upper divertor deposition rate is reduced by \((26/70)^2\). The \ltx evaporator heater was rated at 80~A and operated at 60~A, corresponding to \(I^2R \simeq 1.4\)~kW for a 0.4~\(\Omega\) heater. Operation at 75~A would increase the heater power to about 2.3~kW, a factor of 1.6 higher. If the deposition rate is scaled linearly with this heater power increase, the expected upper divertor lithium deposition rate is about 4.5~nm/min. A 25--100~nm coating would therefore require about 5.5--22~min of active deposition. Including the measured heating time, scaled from the 60~A trace, and cooldown to below 250$^\circ$C gives an estimated NSTX-U evaporation cycle time of about 14--31~min for 25--100~nm coatings. Cooling to 180$^\circ$C instead would increase the upper end to about 34~min. This estimate is intentionally approximate because the lithium evaporation rate is set by vapor pressure and basket temperature, not directly by heater power, but it gives the relevant engineering scale for between shot operation or operation during a run day. The cooldown part of this extrapolation depends on remaining lithium inventory, basket and evaporator head thermal mass, and conductive heat paths after the heater is turned off. The liquid lithium dropper enables precise control of the evaporator head inventory, which was about 3~g in the \ltx tests. The NSTX-U implementation can therefore be optimized by limiting the loaded inventory.

\begin{figure}[htbp]
\centering
\includegraphics[width=0.45\columnwidth]{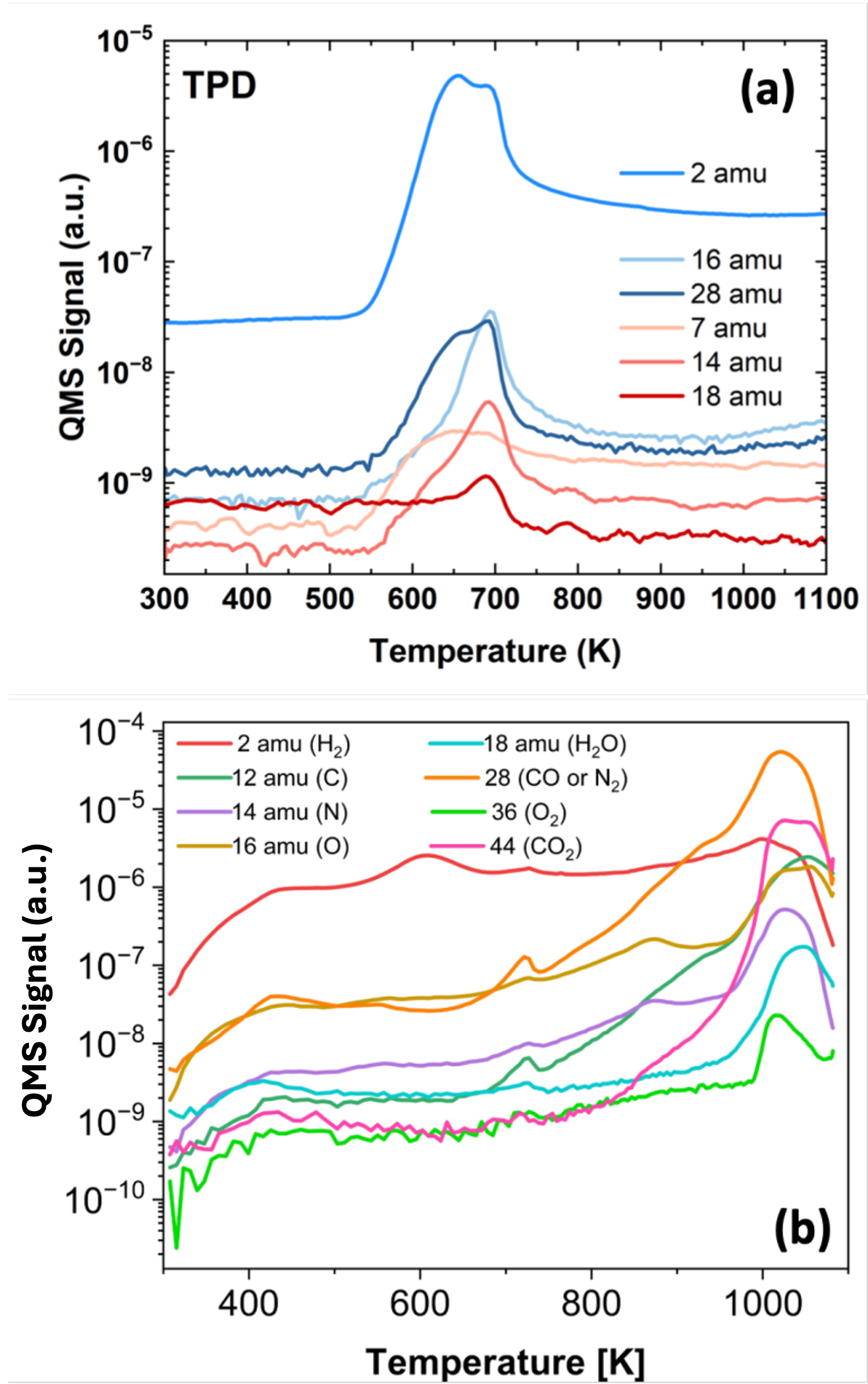}
\caption{\label{fig:mark-iv-tpd}Temperature programmed desorption comparison following lithium deposition on the SEP witness sample. Both panels report uncalibrated QMS signal in arbitrary units, so amplitudes are not quantitatively comparable between panels. (a) Deposition from the evaporator loaded in vacuo using the liquid dropper is dominated by the 2~amu channel, with weaker impurity bearing mass channels. (b) Deposition from the manually loaded solid lithium evaporator shows H$_2$ together with O, C, N, H$_2$O, CO/N$_2$, O$_2$, and CO$_2$ associated mass channels.}
\end{figure}

Testing in DART and \ltx demonstrated that the dropper could deliver molten lithium to the felt lined basket and that the loaded source could be operated without the solid lithium transfer step. The evaporator tested on DART was moved to \ltx and was used routinely during its last campaign. \ltx has a dedicated Sample Exposure Probe (SEP) \cite{maan_sep}, which has been used for in vacuo surface analysis and plasma exposure studies \cite{Maan_PPCF_2020,Maan_thesis_2020,Maan_IEEE_2020,Ostrowski_thesis_2024,lopezperez2026sps-tungsten-gdc}. The SEP is a vacuum transfer probe with a witness sample that can be inserted at the low field side midplane of \ltx for plasma and lithium exposure. After exposure, the witness sample can be pulled back for temperature programmed desorption (TPD), or the probe can be detached from the torus and wheeled out of the test cell to a dedicated surface analysis station in another laboratory while maintaining UHV conditions. TPD is a surface science technique in which a sample is heated in vacuum at a controlled ramp rate. Gases desorbing from the sample are measured, typically using a quadrupole mass spectrometer. The desorption rate and desorption temperature are related to the activation energy of the bound species. In tokamaks, TPD has been used to diagnose hydrogen and impurity retention in PFCs \cite{mapp}. Figure~\ref{fig:mark-iv-tpd} shows a TPD comparison for two lithium evaporations on the SEP: one from a Mark-II evaporator with a solid lithium load \cite{Ostrowski_thesis_2024} and one from an in vacuo liquid lithium load using the dropper. The comparison is used here as an engineering screening measurement rather than as an absolute inventory measurement. Because the QMS response factors were not absolutely calibrated, both vertical axes are reported in arbitrary units and only qualitative spectral differences are interpreted. The case loaded using the liquid dropper in Fig.~\ref{fig:mark-iv-tpd}(a) is dominated by the 2~amu channel and has weaker impurity bearing channels. The case loaded with solid lithium in Fig.~\ref{fig:mark-iv-tpd}(b) shows H$_2$ together with clear O, C, N, H$_2$O, CO/N$_2$, O$_2$, and CO$_2$ associated channels. Within the limitations of this qualitative comparison, the weaker impurity bearing response after loading with the liquid dropper is consistent with reduced contamination introduced during loading.

\subsection{\label{sec:nstxu}NSTX-U f-LITER design}

\begin{figure}[htbp]
\centering
\includegraphics[width=0.9\columnwidth]{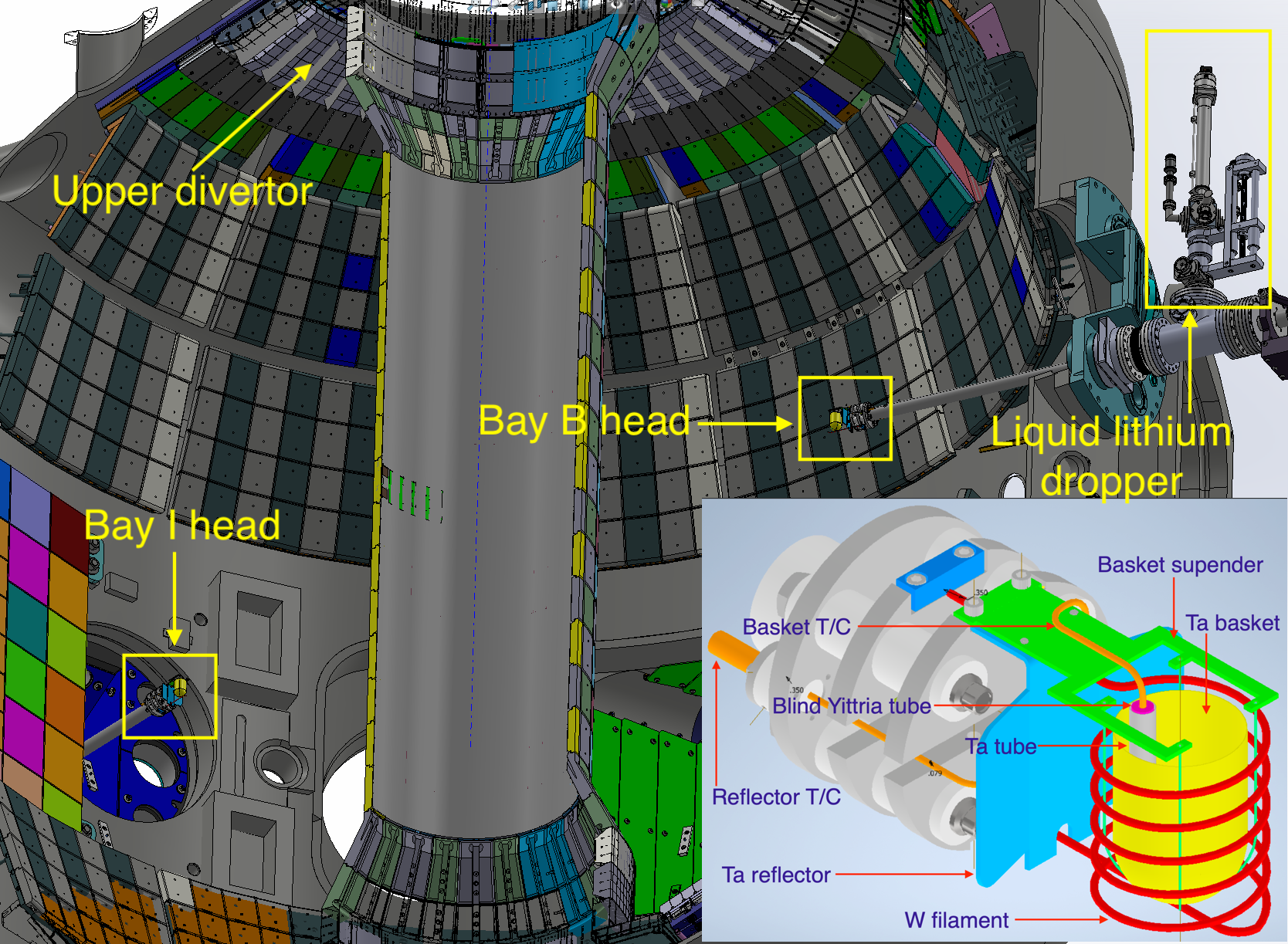}
\caption{\label{fig:nstx-u-integration}NSTX-U vacuum vessel integration concept for the f-LITER system. Two midplane evaporator heads are planned, one at toroidal Bay B and one at toroidal Bay I. The CAD section view shows both heads in the fully inserted position, with the Bay B liquid lithium dropper mounted outside the vessel on the probe assembly. The inset shows the detachable f-LITER head, including the tantalum basket, tungsten filament, tantalum reflector, yttria insulating tube, tantalum tube, basket suspender, and dedicated thermocouples for the basket and reflector.}
\end{figure}

The Mark-II evaporator head was built around an evaporator carriage brazed to two copper conductors that ran back through the bellows to a pair of electrical feedthroughs. This layout came from the earlier LTX yttria crucible evaporators that required a few hundred amperes of current. The flash evaporators can run with \(<100\)~A, for which 0.25~in. diameter oxygen-free high-conductivity copper rods are sufficient. The NSTX-U f-LITER implementation can therefore be simplified by making the head detachable. This has two advantages. First, if a head fails, the probe can be serviced without removing the bellows from the test cell. Second, different heads with different reflector configurations can be used as the coating requirements evolve. Figure~\ref{fig:nstx-u-integration} shows the NSTX-U vessel integration concept and detachable f-LITER head. The design keeps the heated basket assembly removable. The basket, filament, reflector, and thermocouples are grouped into a single serviceable head. This preserves the low thermal mass Mark-II architecture while making replacement and maintenance more practical for NSTX-U.

A liquid lithium dropper loads the head in vacuum while the probe is parked outside the vessel behind the torus interface valve. A 2.75~in. CF manual gate valve separates the dropper from the evaporator, so an empty dropper can be swapped without venting the evaporator. The probe bellows must provide enough stroke to move between park and operating positions. The same architecture is compatible with remote actuation of the probe drive on a shot cycle. NSTX-U requires at least two such probes, placed as close to 180$^\circ$ apart as possible. The reflector must insert past the plane joining the plasma facing surfaces, on the low field side, above and below the midplane so the intended coating region is reached. At the torus interface a 6~in. or 8~in. flange with a pneumatic gate valve is preferable for integration and service. The parked probe also needs its own pumping path so the bellows volume does not build up pressure when the torus interface valve is closed before a shot. A short bellows nipple between the probe and the torus interface valve will be included to mechanically decouple the assembly from the torus.

\section{\label{sec:discussion}Summary and conclusion}

Lithium conditioning was used on LTX and \ltx to modify the plasma boundary condition by covering plasma facing components made from high atomic number materials with low atomic number lithium. When the \ltx shell was well coated, the device accessed low recycling discharges with lower density, reduced neutral influx, stronger particle pumping, and reduced gradient electron temperature profiles. The evaporator development path described here was therefore driven by an operational technology requirement: lithium had to be delivered quickly, repeatably, and to the relevant plasma facing surfaces. The original yttria crucible and shell pool methods could coat large areas, but they imposed pumpout, cooldown, or shell heating delays of several hours that exposed fresh lithium to residual gases. Mark-I flash evaporators reduced the evaporation cycle to minutes and made rapid coating practical. Coverage modeling and shell photographs then showed that geometric shadowing left part of the high field side limiter undercoated. Mark-Ia improved that coverage with a less obstructed carriage and a tantalum reflector. Mark-II increased lithium inventory and suppressed dripping with a felt metal liner, but it still required manual solid lithium loading and repeated vents. The remaining limitation was not evaporation from the basket, but the ability to reload and refresh lithium without interrupting operations or adding impurities.

These results define the NSTX-U requirements for a lithium evaporator based on \ltx. The evaporator must coat surfaces relevant to double null operation, preserve the freshest possible lithium state, avoid wasting lithium on shutters or parked structures, and minimize impurity pickup during loading. It must also be serviceable and compatible with remotely actuated between shot operation. The f-LITER described here, which faces upward and is fed by a dropper, is the direct response to those requirements. Its midplane geometry addresses the coating footprint limitation of the downward facing NSTX LITERs. Its liquid lithium dropper addresses the solid lithium loading limitation of the LTX lineage, and the detachable head provides a practical maintenance path for NSTX-U operation. DART and \ltx testing established that 2--3~g dropper loads could wet the felt lined basket, avoid a free lithium molten pool that could potentially lead to drips, and support 100~nm scale evaporation cycles in minutes. TPD measurements showed reduced impurity release after including the liquid lithium dropper to reload the evaporator head in vacuo. Further qualifications that focus on deposition footprint, evaporation rate calibration, dropper reliability, diagnostic window exposure, and integration with the NSTX-U shot cycle are underway and will be completed soon. With those tests completed, f-LITER would provide NSTX-U with a practical route to reproducible lithium conditioning to enable high performance double null operation. In practical terms, it would enable NSTX-U to refresh lithium on upper divertor and center column PFCs without the lower divertor footprint, shutter interception, and refill interventions of the legacy LITER system. A similar evaporator architecture will also be deployed on the ST40 spherical tokamak as part of the Lithium Evaporations to Advance PFCs in ST40 (LEAPS) project \cite{leaps2025}.

The f-LITER architecture described here is the final baseline design intended for installation on NSTX-U, subject to completion of the ongoing qualification tests. Operational experience may motivate later modifications, but those refinements will build from this installation baseline rather than from a separate set of conceptual design recommendations.
\section{\label{sec:ack}Acknowledgement}

The authors acknowledge support from the entire past and present \ltx team, especially J. Armeli, D. Corl, E. Merino, F. Rabanales, and P. Sloboda, who helped assemble and commission the \ltx evaporators. This material is based on work supported by the U.S. Department of Energy, Office of Science, Office of Fusion Energy Sciences, under contract No. DE-AC02-09CH11466.

\section{\label{sec:data}Data availability}

The data that support the findings of this study, including evaporator development records, coverage modeling outputs, representative surface analysis results, and analysis scripts, reside in Princeton Data Commons \cite{maan_anurag_2026}. Princeton Data Commons provides data publication, preservation, discoverability, and DOI assignment through Princeton Research Data Service.

%\nocite{*}
\bibliography{aipsamp}% Produces the bibliography via BibTeX.

\end{document}